\documentclass[aps,prb,nopacs,10pt,a4paper,nofootinbib,twocolumn]{revtex4}
\usepackage[utf8]{inputenc}
\usepackage{graphicx,amsmath}
\usepackage{amssymb}
\usepackage{amsthm}
\usepackage{slashed}
\usepackage{braket}
\usepackage{bbold}
\usepackage{hyperref}
\usepackage[caption=false,singlelinecheck=off]{subfig}
\usepackage{booktabs}
\usepackage{blindtext}

\begin{document}
\title{Extended Majorana zero modes in a topological superconducting-normal T-junction}
\author{Christian Sp\aa nsl\"{a}tt}
\email{christian.spanslatt@fysik.su.se}
\author{Eddy Ardonne}
\affiliation{Department of Physics, Stockholm University, SE-106 91 Stockholm, Sweden}
\date{\today}
\begin{abstract}
We investigate the sub gap properties of a three terminal Josephson T-junction composed of topologically superconducting wires connected by a normal metal region. This system naturally hosts zero energy Andreev bound states which are of self-conjugate Majorana nature and we show that they are, in contrast to ordinary Majorana zero modes, spatially extended in the normal metal region. If the T-junction respects time-reversal symmetry, we show that a zero mode is distributed only in two out of three arms in the junction and tuning the superconducting phases allows for transfer of the mode between the junction arms. We further provide tunneling conductance calculations showing that these features can be detected in experiments. Our findings suggest an experimental platform for studying the nature of spatially extended Majorana zero modes.
\end{abstract}
\maketitle

\section{Introduction}
\label{sec:Introduction}
Majorana zero modes (MZMs) - particle-hole symmetric zero energy excitations - in solid state devices have attracted much attention in contemporary condensed matter physics. This increased research is partially driven by the search for systems with non-Abelian statistics, which is expected to realize topological quantum computation, and also since the MZMs provide a signature of a novel phase of matter - topological superconductivity \cite{Kitaev:2007gb,alicea2012new,nayak2008non}.

One system predicted to host MZMs is a one-dimensional (1D) semi-conducting nanowire, such as $InAs$ or $InSb$, with strong spin-orbit coupling and large $g$-factors, in proximity with an $s$-wave superconductor (SC) in an external magnetic field \cite{OppenTSC,SauTSC}. Above a critical magnetic field strength, the system is effectively a spinless $p$-wave SC and MZMs are expected to appear on the edges of the wire. The zero energy and particle-hole properties of the MZMs are further predicted to give rise to a robust quantized tunneling conductance of $2e^2/h$ at zero voltage bias due to perfect Andreev reflection \cite{flensberg2010tunneling,fidkowski2012universal,DiezTunneling,SarmaAnalytical}. Several experiments \cite{LeoMaj,LundExp,das2012zero} have reported zero bias peaks (ZBPs) in this type of wires, although the quantization of the conductance has so far not been observed and alternative explanations for robust ZBPs, not related to MZMs, have been proposed \cite{AltlandZeroBias,LeeZeroBias}.

More generally, a 1D wire hosting MZMs is but one manifestation of the various phases of matter predicted by
the recently established periodic table of topological superconductors and insulators \cite{AltlandZirnbauer,Schnyder2008,KitaevPeriodic,RyuLudwig,fulga2012scattering}. In this table, gapped and free fermion systems are classified according to their anti-unitary symmetries and spatial dimension. For a given system, one may construct a mathematical quantity, a topological invariant, associated with the band structure of the bulk, and its value determines whether the system is in a topologically trivial or non-trivial phase. The crucial property of this entity is that it can not change unless the gap closes provided certain symmetries remain intact. 

A non-trivial bulk topology of such gapped phases of matter is expected to give rise to various exotic boundary modes such as MZMs for finite systems. A short ``bulk-boundary'' argument for this statement is that any topologically non-trivial system must change its topological invariant when bordering a topologically distinct domain, for instance the trivial vacuum. Since the invariant can not change without a closing of the gap, gapless modes appear at the system boundary.

It is however not entirely clear what happens to boundary modes in contact with gapless phases of matter. For those, there are no topological invariants defined and one can not use arguments such as the one given above. In the context of localized edge MZMs, there have been some investigations what happens to the edge modes of a topological SC coupled to a finite normal gapless metal (NM)\cite{fidkowski2012universal,stanev2014quasiclassical,hui2014generalized,valentini2014andreev,LossLeaking,
ChevalierLeaking}. The conclusion is that the Majorana mode is exponentially localized in the SC region but extends into the whole NM with a uniform density, while keeping its zero energy and particle-hole symmetric properties. To highlight this feature, we shall refer to such modes as ``extended'' Majorana zero modes (EMZMs). The density of states contribution of the mode decreases as $1/L_N$, where $L_N$ is the NM length. For $L_N$ on the order of the SC coherence length, a gap is induced in the NM due to the SC proximity effect and there is a finite energy gap between the EMZM and neighbouring low energy modes.

In this work, we study a system of three 1D nanowires, of the type mentioned above, constituting the arms in a T-junction, see Figs. \ref{fig:ScatteringSetup} and \ref{fig:TJunctionSetup}. Each wire is driven into an effectively spinless regime by Rashba spin-orbit coupling and an external magnetic field, and the outer regions of the system are SC by proximity. This configuration of wires is effectively a three-terminal Josephson junction of spinless $p$-wave SCs and we assume the SC phases to be externally controllable. Similar setups were previously investigated in the context of transport or braiding \cite{alicea2011non,HalperinJunctions,hyart-qc,ChineseT,Weithofer,IndianT} of MZMs, while in this work we are interested in the nature of EMZMs. We will therefore consider T-junctions in the long junction limit, so that we can really distinguish EMZMs from ``ordinary'' MZMs.

In Sec. \ref{sec:A T-junction of 1D topological superconductors}, by using an analytical scattering matrix approach, we derive three key features of the T-junction. \textit{(i)} There is always at least one EMZM located in the NM region regardless of the SC phases, \textit{(ii)} a single EMZM's spatial distribution is shown to strongly depend on the SC phases, suggesting protocols for transferring them between the arms of the junction by tuning the SC phases, and \textit{(iii)} if the system respects a ``pseudo'' time-reversal symmetry (PTRS), there can be three EMZMs located in the NM region. 

In Sec. \ref{sec:Numerical Calculations}, we confirm these findings numerically with a tight-binding model and with scattering matrix methods we show how the results can be probed experimentally by tunneling spectroscopy. We briefly discuss experimental aspects in Sec. \ref{sec:Experimental Aspects} and in Sec. \ref{sec:Conclusions} we end with a summary and some concluding remarks. 

\section{A T-junction of 1D topological superconductors} 
\label{sec:A T-junction of 1D topological superconductors}
\begin{figure}[t]
\includegraphics[width=1.0\columnwidth]{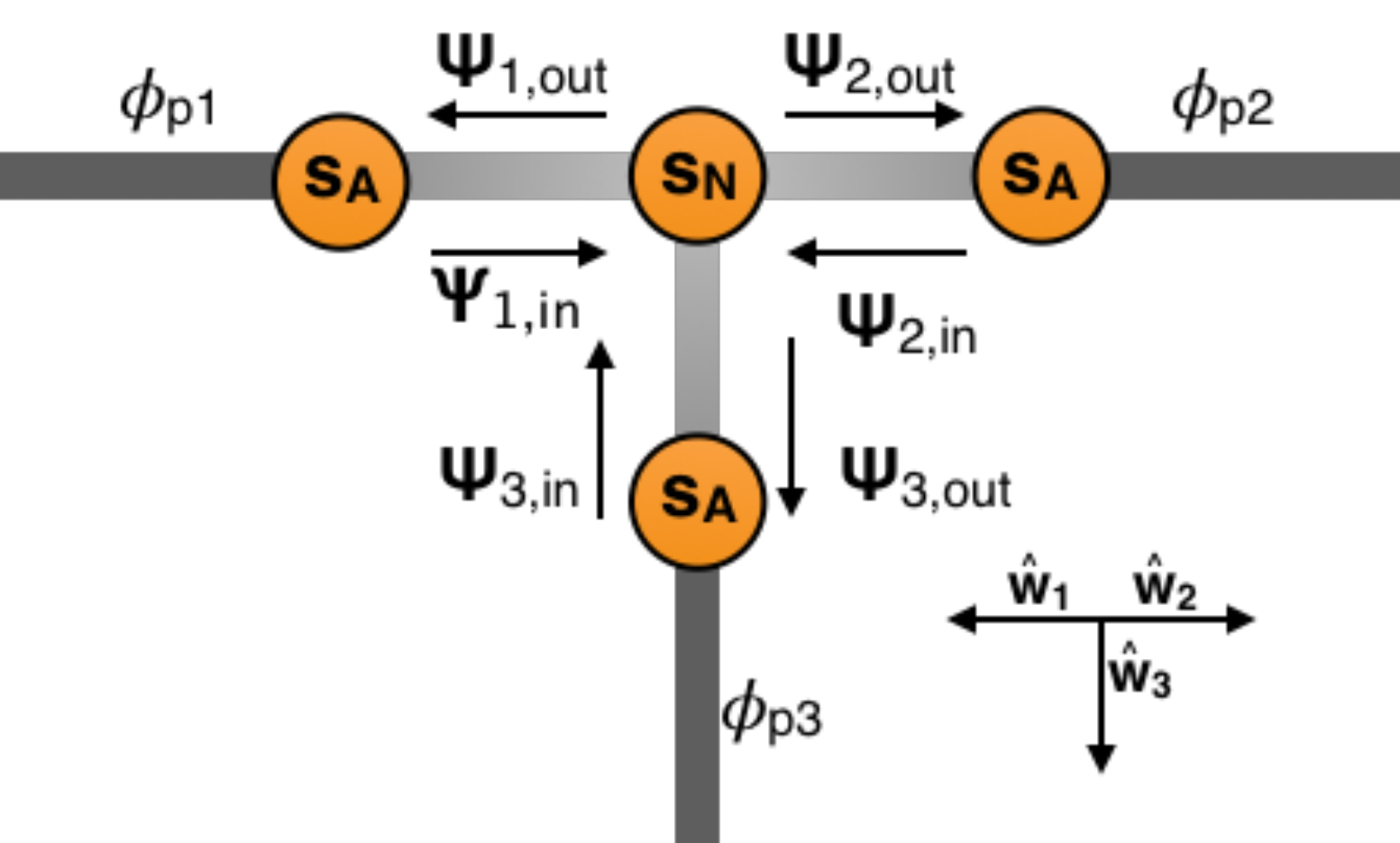}
\caption{Schematics of the scattering approach to the T-junction Josephson setup. Electron and hole scattering states $\Psi_{j,\text{in/out}}$ in the central normal metal region (light grey) are connected by the normal scattering matrix $s_N$. For energies below the superconducting gap, the scattering states leaving the central region Andreev reflect at the three outer topological superconductors (dark grey) with phases $\phi_{p1}$, $\phi_{p2}$, and $\phi_{p3}$ for left, right, and bottom arms respectively. The Andreev processes are described by a single scattering matrix $s_A$. Each arm's $w$-coordinate is chosen to increase from the origin - the central connection point. The relative length between normal and superconducting segments is not drawn to scale.}
\label{fig:ScatteringSetup}
\end{figure}

We consider a three-terminal Josephson junction setup, illustrated in Fig. \ref{fig:ScatteringSetup}, where three spin-less $p$-wave SC wires are connected by spin-less NM wires forming an SNS T-junction. The SCs are assumed to have the same gap $|\Delta_p|$ but may have different SC phases $\phi_{p1}$, $\phi_{p2}$, and $\phi_{p3}$ respectively. 

This system is described by a Hamiltonian, $\mathcal{H}$, fullfilling the intrinsic anti-unitary particle-hole symmetry (PHS) $\mathcal{P}\mathcal{H}\mathcal{P}^{-1}=-\mathcal{H}$ with $\mathcal{P}^2=+1$. In terms of topological classification, this Hamiltonian generally belongs to symmetry class $\mathcal{D}$ \cite{AltlandZirnbauer}. We choose a basis where $\mathcal{P}=\tau_x \mathcal{K}$. The
Pauli matrices  $\tau_x$, $\tau_y$ and $\tau_z$ act in particle-hole space and $\mathcal{K}$ denotes complex conjugation. Furthermore, the system  obeys PTRS if there is an anti-unitary operator $\mathcal{T}$ such that $\mathcal{T}\mathcal{H}\mathcal{T}^{-1}=\mathcal{H}$ with $\mathcal{T}^2=+1$. With PTRS in addition to the PHS described above, the system belongs to class $\mathcal{BDI}$. We choose our basis such that $\mathcal{T}=\mathcal{K}$. 

We start by investigating the low energy features of this setup using a scattering matrix approach. In this way, we don't have to worry about any microscopical details. In particular, the results we find below are still valid in the presence of weak disorder respecting the symmetry classes, assuming the disorder does not close the gap. The symmetry constraints of $\mathcal{H}$ are straightforwardly implemented as described next. 

\subsection{Scattering approach and bound state equation}
\label{sec: ScatteringTheory}
In the NM region, each arm is described by a Bogoliubov-de-Gennes (BdG) Hamiltonian
\begin{equation}
\label{eq:BdGHam1}
 \mathcal{H}_{BdG}(w) = \left(-\frac{\partial_w^2}{2m}-\mu \right)\tau_z, 
\end{equation}
where $w$ is the direction along the wire, $m$ is the effective mass, and $\mu$ is the chemical potential. Throughout this section, we set $\hbar = e = 1$. We assume that the chemical potential and effective mass is the same for all three arms. The flux-normalized \cite{DattaBook} free electron and hole solutions of Eq. \eqref{eq:BdGHam1} are given by
\begin{subequations} 
\label{eq:BdGSolTot}
\begin{align}
\psi^e_\text{in} &=\frac{1}{\sqrt{k_e}} \begin{pmatrix}
1 \\
0
\end{pmatrix} e^{-ik_e w} \label{eq:BdGSol1}\\
\psi^h_\text{in} &= \frac{1}{\sqrt{k_h}}\begin{pmatrix}
0 \\
1
\end{pmatrix} e^{-ik_h w} \label{eq:BdGSol2} \\
\psi^e_\text{out} &=\frac{1}{\sqrt{k_e}} \begin{pmatrix}
1 \\
0
\end{pmatrix} e^{+ik_e w} \label{eq:BdGSol3}\\
\psi^h_\text{out} &=\frac{1}{\sqrt{k_h}} \begin{pmatrix}
0 \\
1
\end{pmatrix} e^{+ik_h w}. \label{eq:BdGSol4}
\end{align}
\end{subequations}
Here, $k_{e,h}=\sqrt{2m(\mu \pm \epsilon)}$ are the wave vectors for electrons and holes respectively and $\epsilon$ is the energy. We use a directional convention where each arm separately has its positive direction pointing away from the central connection point taken to be $w=0$. The subscript ``in/out'' then refers to incoming or outgoing states with respect to this point. 

We use the states in Eq. \eqref{eq:BdGSolTot} as our scattering basis $\Psi_\text{in/out} = (\Psi^e, \Psi^h)^T_\text{in/out}=(\psi^e_1,\psi^e_2,\psi^e_3, \psi^h_1,\psi^h_2,\psi^h_3)^T_\text{in/out}$.

With this construction, the coupling between the three NM arms is fully described by the scattering matrix equation
\begin{equation}
\label{eq:NScatt}
\Psi_\text{out} = s_N(\epsilon) \Psi_\text{in}.
\end{equation}

Generally, the symmetry relations obeyed by the complete system Hamiltonian $\mathcal{H}$ are carried over to the scattering matrix. In the basis we use, any scattering matrix $s_\mathcal{H}(\epsilon)$ relating eigenstates of $\mathcal{H}$ has to obey the relations \cite{fulga2012scattering,BeenakkerMajorana}
\begin{subequations}
\begin{align}
 &s_\mathcal{H}(\epsilon) = \tau_x s_\mathcal{H}^*(-\epsilon)\tau_x, \quad & \text{in classes $\mathcal{D}$ and $\mathcal{BDI}$} \label{eq:DS}\\ 
&s_\mathcal{H}(\epsilon) = s_\mathcal{H}^T(\epsilon), \quad &\text{in class $\mathcal{BDI}$ only.} \label{eq:BDI}
\end{align}
\end{subequations}
In the NM region, scattering does not mix electrons and holes and the scattering matrix is block-diagonal in particle-hole space:
\begin{equation}
\label{eq:SN}
s_N(\epsilon) = \begin{pmatrix}
s(\epsilon) & 0 \\
0 & s^*(-\epsilon)
\end{pmatrix}.
\end{equation}
We assume that the wires only have one single active channel each, so that $s(\epsilon)$ is a $3\times3$ unitary matrix that connects electron states between the different arms in the junction. The scattering matrix relating hole states is given by $s^*(-\epsilon)$ as follows from Eq. \eqref{eq:DS}. The implementation of additional channels or arms is straight-forward but is not considered in this work. 

If the SCs are in their topological regime, that is hosting edge MZMs, the Andreev scattering processes at the NS interfaces are described by the following scattering matrix equation\cite{BrouwerZeroBias,Lee2009,FulgaQuantumNumber,AltlandZeroBias,SarmaAnalytical}
\begin{equation}
\label{eq:AScatt}
\Psi_\text{in} = s_A(\epsilon) \Psi_\text{out},
\end{equation}
with 
\begin{equation}
\label{eq:SA}
s_A(\epsilon) = \alpha(\epsilon) \begin{pmatrix}
  0 & -r_A^* \\
r_A & 0
\end{pmatrix}.
\end{equation}
Here, $\alpha(\epsilon) = e^{-i\arccos(\epsilon/|\Delta_p|)}$ is the usual phase matching factor in the regime $\epsilon \ll |\Delta_p|$ \cite{Andreev,BeenakkerAndreev,beamsplitter} (we refer to Appendix \ref{sec:Symmetries of the reflection matrix} for a brief discussion of scattering onto a spin-less 1D topological SC).

The matrix $s_A(\epsilon)$ is unitary if we assume no single particle transmission into the SCs which is reasonable in the subgap regime. We emphasize the relative sign between the off-diagonal blocks of $s_A(\epsilon)$ which indicates the $p$-wave nature of the pairing. The Andreev reflection matrix $r_A$ is given by
\begin{equation}
\label{eq:rA}
r_A =i\begin{pmatrix}
e^{i \phi_{p1}} & 0 &0 \\
0& e^{i \phi_{p2}} & 0 \\
0&0& e^{i \phi_{p3}}
\end{pmatrix},
\end{equation}
and encodes the phase information acquired by Andreev reflected electrons and holes. With this expression, we have assumed perfect Andreev reflection, the Andreev approximation\cite{Andreev,BeenakkerAndreev}, which holds exactly for $\epsilon=0$. From Eqns. \eqref{eq:BDI} and \eqref{eq:rA}, we note that PTRS can only be present if each SC phase takes values $\phi_{pj}=n\pi$, where $n$ is an integer.

Electrons and holes that scatter in the T-junction can form Andreev bound states (ABS) due to constructive interference of periodic scattering paths. In the setup considered here, with effectively spinless particles, an ABS at the Fermi level, $\epsilon=0$, satisfies the Majorana criterion ($\psi_0^\dagger = \psi_0$) but in contrast to a localized MZM, this mode is spatially extended across the arms in the junction. It is therefore of the EMZM type described in the introduction.   

With Eqns. \eqref{eq:NScatt} and \eqref{eq:AScatt}, the condition for ABS in the T-junction is given by
\begin{equation}
\label{eq:BSCond}
s_A(\epsilon) s_N(\epsilon) \Psi_\text{in} = \Psi_\text{in}.
\end{equation}
This equation serves as our starting point to examine the conditions for having EMZMs in the T-junction and how these modes are spatially located. 

\subsection{Existence and location of EMZMs}
\label{sec:Existence and location of EMZMs}
We now show \textit{(i)} Eq. \eqref{eq:BSCond} has always at least one solution for $\epsilon=0$, \textit{(ii)} a single $\epsilon=0$ bound state is always located in only two of the three arms in the junction if the NM region respects PTRS and two of the SC phases are equal, and \textit{(iii)} if the total system is in class $\mathcal{BDI}$, there are three $\epsilon=0$ solutions only if the three SC phases are equal. Otherwise, there is only one solution. 

Following Ref. \onlinecite{vanHeckTriJunction}, we rephrase Eq. \eqref{eq:BSCond} as
\begin{equation}
\label{eq:BSCond2}
\begin{pmatrix}
s^\dagger(\epsilon) & 0 \\
0 & s^T(-\epsilon)
\end{pmatrix} 
\begin{pmatrix}
0 & r_A^* \\
-r_A & 0
\end{pmatrix} 
\Psi_\text{in}=
\alpha(\epsilon)\Psi_\text{in},
\end{equation}
where we have used Eqns. \eqref{eq:SN} and\eqref{eq:SA}. Adding Eq. \eqref{eq:BSCond2} to its inverse, which conveniently maps $\alpha(\epsilon)$ to $2\epsilon/|\Delta_p|$, gives a new eigenvalue equation
\begin{equation}
\label{eq:BSCond3}
\begin{pmatrix}
0 & A^\dag(\epsilon) \\
A(\epsilon) & 0
\end{pmatrix} 
\Psi_\text{in}=\frac{\epsilon}{|\Delta_p|}\Psi_\text{in},
\end{equation}
with
\begin{equation}
\label{eq:DefA}
A(\epsilon) \equiv \frac{1}{2}(r_A s(\epsilon)-s^T(-\epsilon)r_A).
\end{equation}
Let us now first investigate the case where the SC phases can take arbitrary values in $[0,2\pi]$ so that the system generally belongs to class $\mathcal{D}$. We seek solutions for $\epsilon=0$ and Eq. \eqref{eq:BSCond3} reduces to
\begin{equation}
\label{eq:A02}
A(0)\Psi^e_\text{in}=0,\; A^\dagger(0)\Psi^h_\text{in}=0
\end{equation}
and the bound state condition becomes
\begin{equation}
\label{eq:DetA}
\text{Det}(A(0)) = 0.
\end{equation}
Now, since $r_A$ is symmetric, $A(0)$ is an odd-dimensional anti-symmetric matrix (see Eq. \eqref{eq:DefA}) and it follows that $\text{Det}(A(0))=\text{Det}(-A^T(0))=-\text{Det}(A(0))=0$. This gives that Eqns. \eqref{eq:BSCond3} and \eqref{eq:DetA} are always satisfied for arbitrary unitary $s_N(0)$ and arbitrary SC phases in $s_A(0)$. Because $\Psi^e_\text{in}$ is a solution of $A(0)\Psi^e_\text{in}=0$, it follows that $\Psi^h_\text{in} = (\Psi^e_\text{in})^*$ is a solution of $A^\dagger(0)\Psi^h_\text{in}=0$.

To derive Eq.~\eqref{eq:BSCond3}, we added Eq. \eqref{eq:BSCond2} to its inverse, which may have introduced additional solutions. We therefore have to check for which combination
$\Psi_\text{in} = (\Psi^e_\text{in}, e^{i \chi} \Psi^h_\text{in})^T$ Eq.~\eqref{eq:BSCond2}, or equivalently, Eq.~\eqref{eq:BSCond}, is satisfied, where $e^{i \chi}$ is a phase factor to be determined.
The two constraints one obtains are each other's complex conjugates, and one 
finds that $e^{i \chi} = -i (\Psi^e_\text{in})^T r_A s(0) \Psi^e_\text{in}$, which is indeed
a phase. Thus one can construct only one solution of
Eq.~\eqref{eq:BSCond} from a solution $\Psi^e_\text{in}$ and $\Psi^h_\text{in}$ of
Eq.~\eqref{eq:BSCond3}; we will make use of this fact again below.

We conclude that there is at least one $\epsilon=0$ bound state, an EMZM, in the junction. This is of course expected because both MZMs and EMZMs can only gap out in pairs due to PHS. For similar results in a slightly different system, see for instance Ref. \onlinecite{MajoranaQSH}. We note that this result generalizes to all junctions, of the type considered here, with an odd number of arms. 

Next, we show that there are cases, in which the zero bound state does not spread out over the
three arms, but resides completely in two arms only. To obtain this result, we
note that the explicit elements of $A(0)$ in general take the form
\begin{equation}
\label{eq:A0Explicit}
A_{mn}(0) \propto t_{mn}e^{i\phi_{pm}}-t_{nm}e^{i\phi_{pn}},
\end{equation}
where the elements of the scattering matrix, $t_{mn}$ are the electron transmission amplitudes from arm $n$ to arm $m$.
If the normal region respects PTRS, we have $t_{nm}=t_{mn}$, by virtue of Eq.~\eqref{eq:BDI}.
If in addition two
of the phases are equal, say $\phi_{p1} = \phi_{p2} \neq \phi_{p3}$, and recalling that $A(0)$ is anti-symmetric,
we find that the only non-zero elements of $A(0)$ are $A_{13}=-A_{31}$ and
$A_{23}=-A_{32}$.

Thus we find that, in this case, the third component of $\Psi^e_\text{in}$ must be zero in order
to satisfy Eq. \eqref{eq:A02}. The same argument applies to the hole-part $\Psi^h_\text{in}$, which
will have a zero in the same position.
As we saw above, these two parts have to be combined in the proper way to obtain a solution of
Eq. \eqref{eq:BSCond}, with the zero elements carrying over. We conclude that in the case that the normal region respects PTRS, and
if two of the phases are equal, there is a single zero energy bound state, that resides completely
in the two arms whose phases are equal.

Finally, we consider the case $\phi_{p1}=\phi_{p2}=\phi_{p3}=0$ (or any phase equal for all arms, because an overall phase corresponds to a global gauge choice), which gives $r_A \propto \mathbb{1}$. If in addition $s_N=s_N^T$, the system belongs to class $\mathcal{BDI}$ and one has $A(0)=0$. Then there
are obviously three solutions $\Psi^e_\text{in}$ of Eq. \eqref{eq:A02} and similarly for
$\Psi^h_\text{in}$. By taking the appropriate combinations, we find three solutions of
Eq. \eqref{eq:BSCond}, so in this case, there are three EMZMs in the junction. We note that in this
case, one can also use Eq. \eqref{eq:BSCond} directly to obtain this result.

Thus, in class $\mathcal{BDI}$ there are three EMZMs if the phases are equal. If one phase is shifted by $\pi$ with respect to the other two, the system remains in $\mathcal{BDI}$, but two EMZMs gap out leaving a single EMZM in the junction as described above. 
 
To understand the physics behind these results it is useful to first analyze Andreev reflection onto a topological SC in NS and SNS junctions. 

For an NS junction, in addition to the phase of the order parameter, $\phi_p$, the Andreev reflection processes $e\rightarrow h$ and $h\rightarrow e$ are phase shifted by exactly $\pi$ for $\epsilon=0$ due to the $p$-wave pairing. This type of phase shift occurs because incoming electrons and outgoing holes (the $e\rightarrow h$ process) with Fermi momentum $p_F$ experience an effective gap $\Delta_p \sim +|\Delta_p|p_F$, while incoming holes and outgoing electrons (the $h\rightarrow e$ process) have momentum $-p_F$ and the experienced gap is $\Delta_p \sim -|\Delta_p|p_F$.
In that sense, for $\epsilon=0$, a $p$-wave SC is analogous to an optical phase-conjugating mirror
(this behaviour is similar for a $d$-wave SC\cite{Tanaka1995}, but in contrast to the case of an $s$-wave SC).
We refer to Ref.~\onlinecite{BeenakkerResistance} for a comparison. 

With this mechanism, an NM connected to a topological SC becomes completely transparent for states at $\epsilon=0$, since any net phase accumulated by an electron-hole-electron or a hole-electron-hole orbit close to the interface becomes zero and multiple paths interfere constructively. In this way, it is clear that a MZM will ``leak out'' from a topological SC into a connected finite NM.

For an SNS junction, similar arguments apply. A phase difference of exactly $\pi$ between the two SCs induces phase-shifts for the Andreev orbits such that the normal region becomes completely transparent at $\epsilon=0$. This behaviour is captured in the $4\pi$-Josephson relation \cite{Kwon2004}, $\epsilon_\text{ABS} = \pm \Delta_p \sqrt{D}\cos(\frac{\Delta \phi_p}{2})$, where $\Delta_p$ is the SC gap, $D$ is the junction transparency, and $\Delta\phi_p$ is the SC phase difference. This relation is straight-forwardly reproduced with a two-terminal version of Eq. \eqref{eq:BSCond} in the short junction limit with PTRS imposed on $s_N(\epsilon)$.

To discuss the T-junction, we first stress that with our directional convention, a zero phase difference between two arms corresponds (somewhat paradoxically) to a physical phase shift of $\pi$. 

As mentioned above, it is clear that the central region must host at least one EMZM. There is one localized MZM at each outer edge of the system and PHS implies that zero modes always appear in pairs. Therefore, at least one zero energy mode must be located in the NM region.

When only two phases are equal and PTRS is imposed on $s_N$, the phase shifted arm is effectively disconnected from the other two at zero energy, see the discussion below Eq. \eqref{eq:A0Explicit}. The two connected wires form a $\pi$-shifted SNS junction with an EMZM while the other two possible arm pair combinations form junctions that can have ABSs albeit not at zero energy.

When all three SC phases are equal, and the system is in class $\mathcal{BDI}$, the wires are effectively disconnected for zero energy, since $A(0)=0$ with these two constraints. Then the argument for NS-junctions above applies separately for each wire and there are three EMZMs in the junction.  Breaking PTRS, either by removing the symmetry constraint on $s_N(0)$ or by slightly shifting one of the phases, causes two EMZMs to hybridize and gap out. 

With these results in mind, we turn to numerical calculations to verify the predictions in a microscopic setting and we show how they can be tested experimentally. 

\section{Numerical Calculations}
\label{sec:Numerical Calculations}
\subsection{Hamiltonian}
\label{sec:Hamiltonian}
To describe the T-junction microscopically, we start with a model for a single semi-conducting nanowire with strong Rashba spin-orbit coupling (RSOC) lying in proximity to a conventional $s$-wave SC in an external magnetic field. Assuming the wire to be thin, so that only one channel is occupied, we use the 1D BdG Hamiltonian $H_\text{NW} = \frac{1}{2}\int dw \mathbf{\Psi}^\dag(w)\mathcal{H}_\text{NW}\mathbf{\Psi}(w)$ with Nambu basis
$\mathbf{\Psi}(w) = \left[\psi^{}_\uparrow (w),\psi^{}_\downarrow (w),\psi^\dag_\uparrow (w),\psi^\dag_\downarrow(w)\right]^T$, where $\psi_\sigma^\dag(w)$ creates an electron with spin $\sigma$ at coordinate $w$ along the wire. We take the wire direction $\mathbf{\hat{w}}$ to lie in the $x$-$y$ plane. With this convention, 
\begin{align}
\label{eq:NanoWire}
&\mathcal{H}_\text{NW} = \begin{pmatrix}
h(p_w) & h_\Delta \\
h^\dagger_\Delta & -h^T(-p_w)
\end{pmatrix},\\
& h(p_w) = \frac{p_w^2}{2m^*}-\mu -\alpha_R \mathbf{p}_w \cdot \boldsymbol{\sigma} \times \mathbf{\hat{z}}+  \mathbf{h} \cdot \boldsymbol{\sigma}, \notag\\
& h_\Delta = |\Delta|e^{-i\phi_s}(-i\sigma_y),\notag
\end{align}

where $\mathbf{p}_w$ is the momentum operator along the wire, $m^*$ is the effective electron mass, $\mu$ is the chemical potential, and $\alpha_R$ is the RSOC strength originating from an internal electrical field pointing in the $\mathbf{\hat{z}}$ direction. The spin-orbit direction is then restricted to lie in the $x$-$y$ plane. Further, $\mathbf{h}\equiv \frac{1}{2}g\mu_B \mathbf{B}$ is the Zeeman field with $g$ the effective g-factor in the wire, $\mu_B$ the Bohr magneton, and $\mathbf{B}$ is the external magnetic field. The proximity induced SC gap is denoted $|\Delta|$ with phase $\phi_s$ inherited directly from the underlying $s$-wave SC. The set of Pauli-matrices $\boldsymbol{\sigma}$ act in spin space.

It has been shown\cite{SauTSC,OppenTSC,HalperinJunctions} that the Hamiltonian \eqref{eq:NanoWire} can be mapped onto a spinless $p$-wave SC model with a topological phase hosting MZMs \cite{Kitaev:2007gb}.
This topological phase occurs when two conditions on the Zeeman field $\mathbf{h}$ are met\cite{CritAngleNum,CritAngleAn}. 
Namely, the full field satisfies the topological criterion $|\mathbf{h}|>h_c \equiv \sqrt{|\Delta|^2+\mu^2}$ (in a finite lattice model, there is an additional upper critical field due to the finite band width, but that field does not play any role in this paper). In addition, the projection $\mathbf{h}_P$ of the Zeeman field onto the $\mathbf{\hat{z}}$-$\mathbf{\hat{w}}$ plane should satisfy $|\mathbf{h}_P| > \sqrt{\mathbf{h}^2-|\Delta|^2}$. This equation sets an upper bound on the component of the Zeeman field pointing in the spin-orbit direction. 

With these conditions in mind, we take the magnetic field to point in the $z$-direction, $\mathbf{h}=\mathbf{h}_P=h \mathbf{\hat{z}}$ for the remainder of the paper. For large magnetic fields, $|\mathbf{h}|\gg \epsilon_\text{SO}$, where the spin-orbit energy $\epsilon_\text{SO}\equiv \alpha^2_R m^*/2\hbar^2$ sets the characteristic energy scale, the effective $p$-wave order parameter is $|\Delta_p|\approx |\alpha \Delta/\mathbf{h}|$. Also, the $p$-wave phase $\phi_p$, depends crucially on the direction of the wire\cite{alicea2011non,HalperinJunctions}. With our coordinate convention, this relation can be written as
\begin{equation}
\label{eq:phases}
\phi_p =  \phi_s + \varphi,
\end{equation}
where $\phi_s$ is again the bulk $s$-wave order parameter and $\varphi = \arccos(\mathbf{\hat{x}}\cdot\mathbf{\hat{w}})$, the angle of the wire with respect to the positive $x$-axis. For a single uniform wire, this extra phase shift is not important due to the gauge freedom to remove any global phase, but for systems with wires coupled at angles it has interesting consequences. 

For instance, two proximity induced wires connected in an ``L-shaped'' Josephson junction exhibit a SC phase difference of $\pi/2$ even if the underlying $s$-wave SCs have the same phase. This observation suggests a generalization of the $\pi$-junction in Ref. \onlinecite{ojanen} (see also Ref.~\onlinecite{Klinovaja2015}) which can be achieved by geometrical means in contrast to arrangements of permanent magnets. Even more interesting, this effect should be manifest in arbitrarily curved wires. For our present purposes, the extra phase shift must be accounted for when modeling the T-junction in order to compare with results from Sec. \ref{sec:A T-junction of 1D topological superconductors}.

\begin{figure}[t]
\includegraphics[width=1.0\columnwidth]{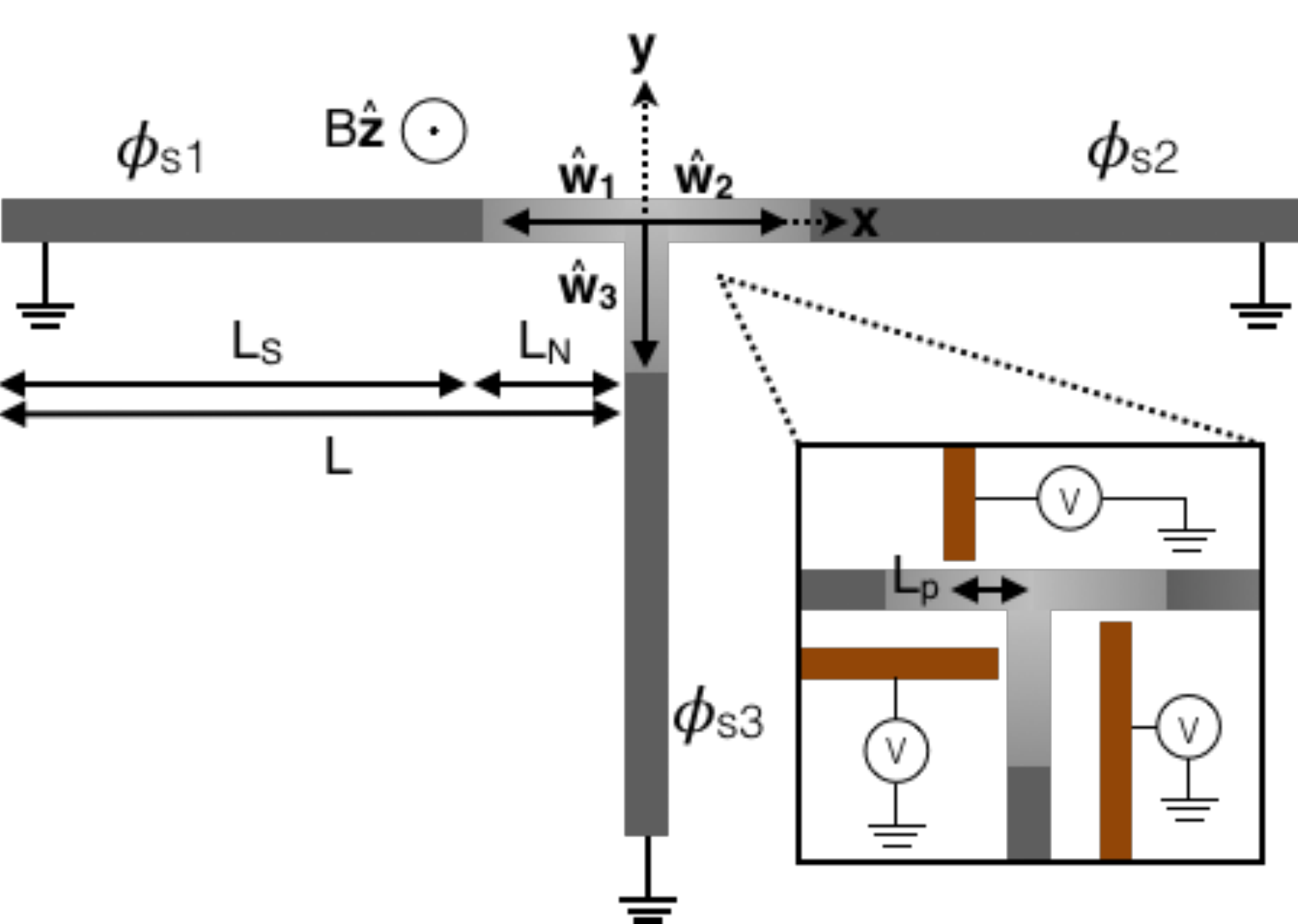}
\caption{Schematics of the T-junction setup. Three nano-wires (light grey regions) in directions $\mathbf{\hat{w}_1}$, $\mathbf{\hat{w}_2}$, and $\mathbf{\hat{w}_3}$ for arm $1$ (left), $2$ (right) and $3$ (bottom) respectively are connected at the origin. The wires are partially (dark grey regions) lying on top of $s$-wave superconductors with phases  $\phi_{s1}$, $\phi_{s2}$, and $\phi_{s3}$ respectively. There is a magnetic field, $B$, in the $\mathbf{\hat{z}}$-direction. The total wire lengths, superconducting segment lengths, and normal metal lengths are denoted $L$, $L_S$, and $L_N$ respectively. \textit{Inset}: a three probe (brown regions) configuration for measuring tunneling conductance with bias voltages $V$.}
\label{fig:TJunctionSetup}
\end{figure}	

We take three wires of the type \eqref{eq:NanoWire}, and discretize them on a lattice with $N=100$ lattice sites per wire. As before, we use a directional convention where wire 1, wire 2, and wire 3 point in directions $\mathbf{\hat{w}_1}=(-1,0,0)$, $\mathbf{\hat{w}_2}=(1,0,0)$, and $\mathbf{\hat{w}_3}=(0,-1,0)$ respectively, see Fig. \ref{fig:TJunctionSetup}. The origin is taken as the connection point and each wire has total length $L=L_N+L_S$, where $L_N$ and $L_S$ denote the normal and SC segment lengths respectively. The discretization introduces the hopping parameter $t\equiv \hbar^2/(2m^*a^2)$ and the Rashba spin flip hopping parameter $\alpha \equiv \alpha_R/(2a)$, where $a\equiv L/N$ is the effective lattice constant. The hopping elements between wires are taken to be pure spin preserving hoppings with $t_c=t/10$. Moreover, we note that with the magnetic field pointing in the $z$-direction, all three wires enter the topological phase simultaneously. We should note that while this is convenient for our purposes, it does limit the extend to which our model calculations can be compared to experiments that use nano-wires with an epitaxially grown superconductor layer\cite{higginbotham}, for which the (perpendicular) critical magnetic field is rather low.

Throughout the remainder of the paper, we use as our unit of energy the spin-orbit energy $\epsilon_\text{SO} \equiv \alpha^2_R m^*/2\hbar^2 \approx 68 \;\mu$eV, where $\alpha_R = 0.2$ eVÅ and $m^* = 0.026m_e$ with $m_e$ being the free electron mass. We also take $g=20$ and $|\Delta|= 2.5\; \epsilon_{SO} \approx 170\; \mu$eV, parameters appropriate for $InAs$ \cite{das2012zero,beamsplitter,FlensbergPhaseTunable}. In terms of the spin-orbit energy, $t\approx 13.4\;\epsilon_\text{SO}$ and $\alpha \approx 3.7 \;\epsilon_\text{SO}$. These choices also define the spin-orbit length  $l_\text{SO} = 2\hbar^2/(m^* \alpha_R) \approx 293$ nm and the SC coherence length \cite{ojanen} $\xi =\epsilon_\text{SO} l_\text{SO}/\Delta \approx 117$ nm.
With our numerical model, we can examine the predictions in Sec. \ref{sec:A T-junction of 1D topological superconductors} and also how they can be verified experimentally by tunneling spectroscopy. The formalism for calculating tunneling conductance is introduced next. 

\subsection{Tunneling spectroscopy}
\label{sec:Tunneling spectroscopy}
Calculations of the tunneling conductance at an NS interface are implemented by the Mahaux-Weidenmüller formula, relating at a given energy the reflection matrix, $r(\epsilon)$, to the Hamiltonian $\mathcal{H}$ by \cite{mahaux1969shell,beenakker1997random}
\begin{equation}
\label{eq:MWFormula}
r(\epsilon) = \mathbb{1} + 2\pi i W^\dag (\mathcal{H}-\epsilon-i\pi W W^\dag)^{-1} W.
\end{equation}
The coupling matrix $W$ is of size $4N\times M$, where $4N$ is the size of the matrix representing $\mathcal{H}$ and $M$ is the total number of lead channels. This matrix contains the coupling elements between the basis states of $\mathcal{H}$ and the modes in the leads. The elements of the matrix $-i\pi W W^\dag$ can be viewed as the lead self-energies, which modify the bare energies and life-times for the particles in the system when leads are attached. 

To attach a single lead at a site $N_p$ with spin- and particle-hole degrees of freedom, we take
\begin{equation}
\label{eq:CouplingMatrix}
W = \sqrt{\lambda}(\vec{v}_p\otimes \mathbb{1}_4)^T,
\end{equation}
where $\vec{v}_p =  (\hdots,0,1_p,0,\hdots)$ is a unit vector of length $N$ representing the site degree of freedom and $\mathbb{1}_4$ is a $4\times4$ unit matrix representing spin and particle-hole degrees of freedom. The coupling between the system and the lead is characterized by the parameter $\lambda$. This construction attaches a lead to site $N_p$ (which corresponds to a distance $L_p=a N_p$ from the origin, see the inset of Fig. \ref{fig:TJunctionSetup}) of the system. 

The reflection matrix in Eq. \eqref{eq:MWFormula} can be divided into particle hole-blocks as
\begin{equation}
\label{eq:refMAt}
 r(\epsilon) =  \begin{pmatrix}
r_{ee}(\epsilon) & r_{eh}(\epsilon) \\
r_{he}(\epsilon) & r_{hh}(\epsilon)
\end{pmatrix},
\end{equation}
where $r_{ee}$ is the reflection amplitude for an incoming electron, $r_{hh}$ is the reflection amplitude for an incoming hole, and $r_{eh}$ and $r_{he}$ are Andreev reflection amplitudes which converts incoming electrons to outgoing holes and vice versa. The tunneling conductance for zero temperature and small bias voltages $V$ is dominated by Andreev processes and is given by \cite{blonder1982transition}
\begin{equation}
\label{eq:ConductanceForm}
G(V) = \frac{2e^2}{h}\text{Tr}\left(r_{eh}^{} (eV) r_{eh}^\dag(eV)\right).
\end{equation}
Using Eqns. \eqref{eq:MWFormula}, \eqref{eq:CouplingMatrix}, \eqref{eq:refMAt}, and  \eqref{eq:ConductanceForm}, we can calculate the subgap tunneling conductance into the T-junction at any site and for any tunneling strength by choosing the coupling matrix $W$ accordingly. 

\subsection{Numerical results}
\label{sec:Numerical results}
In this section we present numerical results for the existence and spatial distribution of EMZMs in the T-junction. Both these entities are shown to depend on the SC phases and can be probed via the tunneling conductance.
This indicates that the predictions in Sec. \ref{sec:A T-junction of 1D topological superconductors} hold and can be verified by tunneling experiments. Throughout this section we choose experimentally relevant lengths\cite{LeoMaj} $L=4.0\;\mu$m, $L_N=0.4\;\mu$m, and $L_S=3.6\;\mu$m for each wire. The distance from the origin to each lead is $L_p=0.12\;\mu$m, see Fig. \ref{fig:TJunctionSetup}. With this choice, we are in the long junction limit since $\xi< L_N$, which is important, because EMZM are only really distinct from ordinary ones in this limit.

We also choose $\mu=0$ for which $h_c=|\Delta|=2.5\;\epsilon_{SO}$. Furthermore, we assume zero temperature.

\subsubsection{Existence and location of EMZMs}
\label{sec:Location of Extended Majorana Zero Modes}
\begin{figure}[t]
\captionsetup[subfigure]{position=top,justification=raggedright}
\subfloat[]{
\includegraphics[width=0.9\columnwidth]{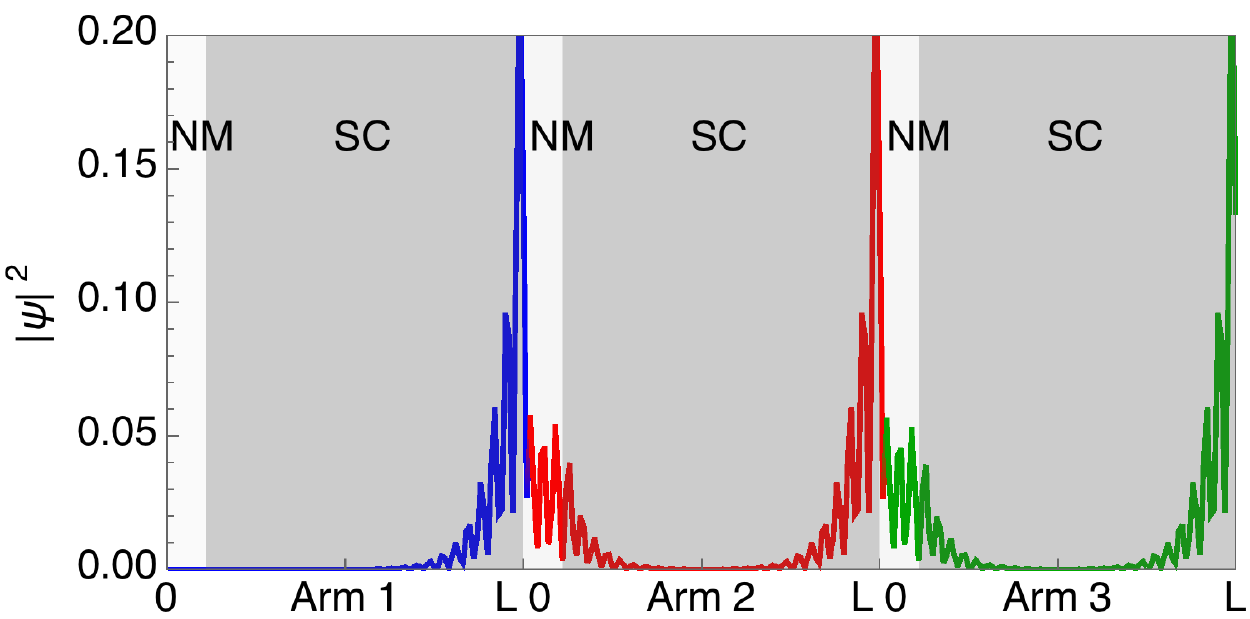}
\label{fig:MajDist1}}

\subfloat[]{
\includegraphics[width =0.9\columnwidth]{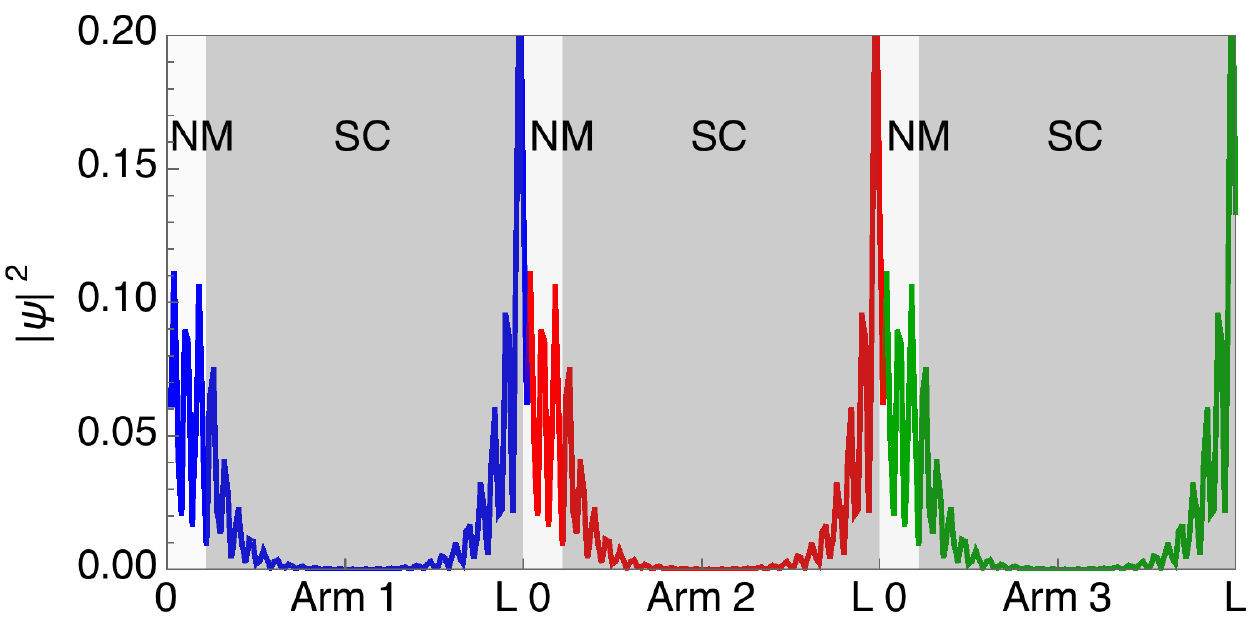}
\label{fig:MajDist2}}
\caption{
Total zero energy mode probability distribution in the T-junction system. The probability weight residing in arms 1, 2, and 3 are plotted in blue, red, and green respectively. The outer superconducting segments (SC) and the normal metal (NM) regions are highlighted with dark and light gray shading respectively. \protect\subref{fig:MajDist1} With phases, $\phi_{s1}=0$, $\phi_{s2}=0$, $\phi_{s3}=\pi/2$, arm 1 is effectively phase shifted with respect to arms 2 and 3. There are four zero energy modes, three exponentially located in the outer edges of the system and one residing in arms 2 and 3 in the central normal region. \protect\subref{fig:MajDist2} With phases, $\phi_{s1}=\pi$, $\phi_{s2}=0$, $\phi_{s3}=\pi/2$, all three arms have effectively the same phase. There are six zero energy modes, three exponentially located in the outer edges of the system and three residing in all three arms in the central normal region.}
\label{fig:MajoranaDistribution}
\end{figure}

We start by investigating how zero energy modes are spatially located in the T-junction. We choose $h=8\;\epsilon_{SO}$ which corresponds to $B_z \approx 0.94$ T. With this choice, the SC segments are in the topological phase. Furthermore, we pick $\phi_{s1}=0$, $\phi_{s2}=0$, and $\phi_{s3}=\pi/2$ which by Eq. \eqref{eq:phases} correspond to $\phi_{p1}=\pi$, $\phi_{p2}=0$ and $\phi_{p3}=0$. We find that the total system hosts four zero modes. The total probability distribution of these modes are displayed in Fig. \ref{fig:MajDist1}. We note that there are three exponentially localized MZMs on the outer edges of the SC wires while there is one EMZM located in arms $2$ and $3$. We checked that the EMZM is always located in the two arms with the same phase. Moreover, this type of spatial distribution is no longer present if complex hoppings are introduced, indicating a breaking of PTRS. Although there is always at least one zero mode in the central region, PTRS breaking makes the mode spread out in all three arms. These results are consistent with Sec. \ref{sec:A T-junction of 1D topological superconductors}. 

Next, we choose $\phi_{s1}=\pi$, $\phi_{s2}=0$, and $\phi_{s3}=\pi/2$, corresponding to $\phi_{p1}=0$, $\phi_{p2}=0$, and $\phi_{p3}=0$. This time we find six zero modes, three exponentially localized MZMs in the outer regions and 3 EMZMs distributed in all three arms in the central region. The total probability distribution of these zero modes is shown in Fig. \ref{fig:MajDist2}. Again, this result agrees with our previous calculations. 

\subsubsection{Tunneling conductance in trivial and topological regimes}
\label{sec:Tunneling conductance in trivial and topological regimes}
Next, we focus on the tunneling conductance. We choose a weak tunneling coupling $\lambda=\;\epsilon_{SO}/4$ and connect tunneling probes at sites $N_p = 3$ of each wire. This site corresponds to a distance $L_p=120$ nm from the origin, see the inset in Fig. \ref{fig:TJunctionSetup}.

\begin{figure}[t]
\captionsetup[subfigure]{position=top,justification=raggedright}
\subfloat[]{
\includegraphics[width=\columnwidth]{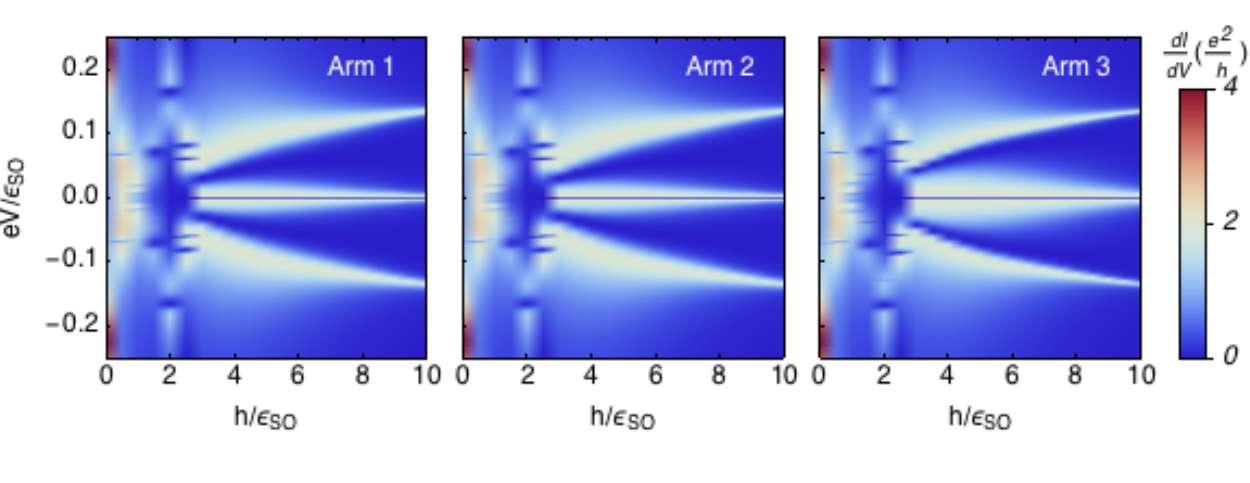}
\label{fig:GvsBa}}
\\[-4ex]
\subfloat[]{
\includegraphics[width =\columnwidth]{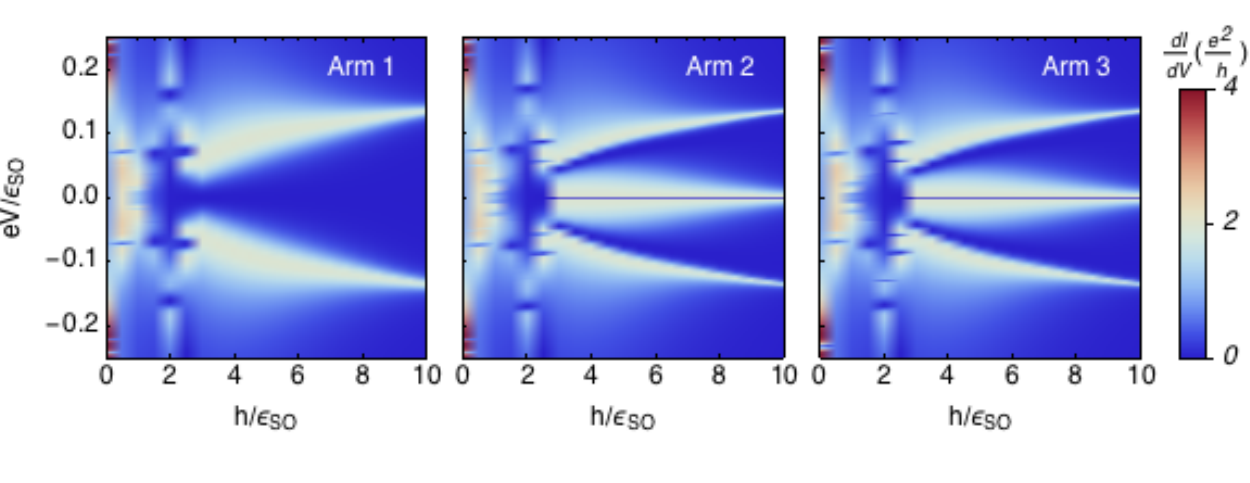}
\label{fig:GvsBb}}
\\[-4ex]
\subfloat[]{
\includegraphics[width =\columnwidth]{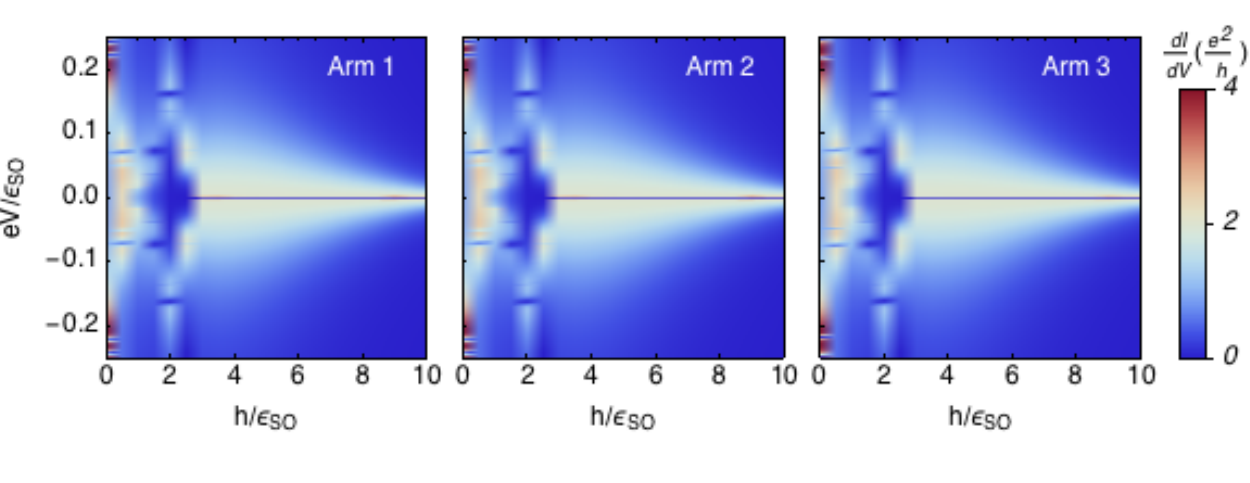}
\label{fig:GvsBc}}
\caption{Differential tunneling conductance for the junction arms at distance $L_p=120$ nm from the origin. \protect\subref{fig:GvsBa} Conductance as a function of bias voltage $V$ and Zeeman field strength $h$ for $\phi_{s1}=\pi/2$, $\phi_{s2}=\pi/2$, $\phi_{s3}=\pi/2$. \protect\subref{fig:GvsBb} Conductance as a function of $V$ and $h$ for $\phi_{s1}=0$, $\phi_{s2}=0$, $\phi_{s3}=\pi/2$. \protect\subref{fig:GvsBc} Conductance as a function of $V$ and $h$ for $\phi_{s1}=\pi$, $\phi_{s2}=0$, $\phi_{s3}=\pi/2$.}
\label{fig:GvsB}
\end{figure}

First, we choose phases $\phi_{s1}=\pi/2$, $\phi_{s2}=\pi/2$, $\phi_{s3}=\pi/2$ and calculate the tunneling conductance for low bias voltages and Zeeman field strengths. From Sec. \ref{sec:A T-junction of 1D topological superconductors}, we expect a single EMZM, manifested by ZBPs, in all three arms when the SC segments are in the topological regime. The result is displayed in Fig. \ref{fig:GvsBa}. We first note that for weak Zeeman fields, the tunneling conductance is higher than $2e^2/h$. This is due to Andreev reflection into the proximitized NM region when the system is in a spin-full regime and the SC segments are non-topological. The Andreev conductance is then non-universal and take values between $0$ and $4e^2/h$ depending on the tunneling coupling\cite{blonder1982transition,fidkowski2012universal}. 

Above the critical field $h_c=2.5$ $\epsilon_\text{SO}$ the system enters a spin-less regime and the SC segments become topological. In this regime ZBPs are observed in each arm. The peaks are not of equal width in all arms, but differ since the EMZM have slightly different weight in the arms.  We further point out that the ZBPs have tiny splittings which we attribute to the non-zero overlap the EMZMs have in a system of finite length\cite{Kitaev:2007gb,flensberg2010tunneling,AguadoSplitting,DasSarmaSplitting,Klinovaja2013}. Furthermore, we note that there are additional low energy modes in all arms. These modes are remnants of additional EMZMs which gap out for the particular phase choice here. For clarity, the conductance spectrum is shown only for bias voltages much smaller than the Zeeman dependent effective gap $|\Delta_p|\approx 1.8$ $\epsilon_\text{SO}$.

We next choose the first set of phases in the previous section, $\phi_{s1}=0$, $\phi_{s2}=0$, $\phi_{s3}=\pi/2$ (which correspond to $\phi_{p1} = \pi$ and $\phi_{p2} = \phi_{p3} = 0$),  for which the results are displayed in Fig. \ref{fig:GvsBb}. We note that there are ZBPs in arms $2$ and $3$ but none in arm 1 which means that the EMZM resides completely in arms 2 and 3, in agreement with the previous section. 
 
We repeat the calculation for the second set of phases from the previous section, $\phi_{s1}=\pi$, $\phi_{s2}=0$, and $\phi_{s3}=\pi/2$ (which correspond to $\phi_{p1} = \phi_{p2} = \phi_{p3} = 0$) and the result is presented in Fig. \ref{fig:GvsBc}. As expected, this time there are ZBPs in each arm. 
With the phase choice made, no EMZMs gap out and all three of them reside in the junction, at zero energy.
Therefore, there are no additional low energy modes. The absence of these low-lying modes indicates that the cases of one and three EMZMs in the junction can be distinguished experimentally, because one can continuously adjust the phases between the two cases. 

These observations lead us to conclude that the EMZMs are manifested by a ZBP in the tunneling conductance in the NM regions of the junction. While the results presented are for a specific point in each arm, we verified that our results do not depend significantly on this choice, as long as the probes are in the NM region. We deduce that in the NM region, the zero modes are indeed ``extended''. The results in this section are consistent with our predictions and moreover, they can be attributed to the topological nature of the SCs.

\subsubsection{Transfer of EMZMs by phase tuning}
\label{Transfer of EMZMs by phase tuning}
The results obtained so far indicate that it should be possible to transfer an EMZM from one arm to another simply by letting the normal region respect PTRS and tuning the $s$-wave SC phases. This feature can be probed by measuring the tunneling conductance of the junction arms when the SC phases are varied. 

\begin{figure}[t]
\captionsetup[subfigure]{position=top,justification=raggedright}
\subfloat[Subfigure 1 list of figures text][]{
\includegraphics[width=0.5\columnwidth]{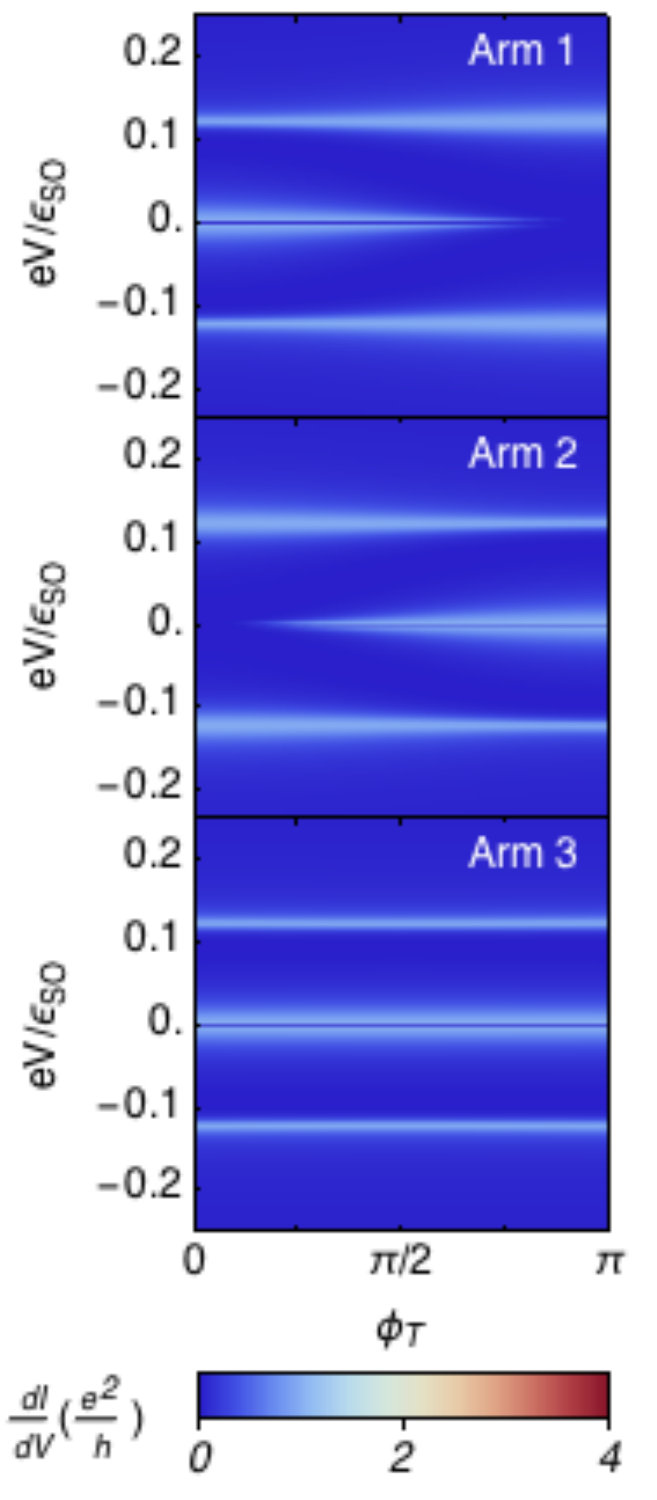}
\label{fig:TransferA}}
\subfloat[Subfigure 2 list of figures text][]{
\includegraphics[width=0.5\columnwidth]{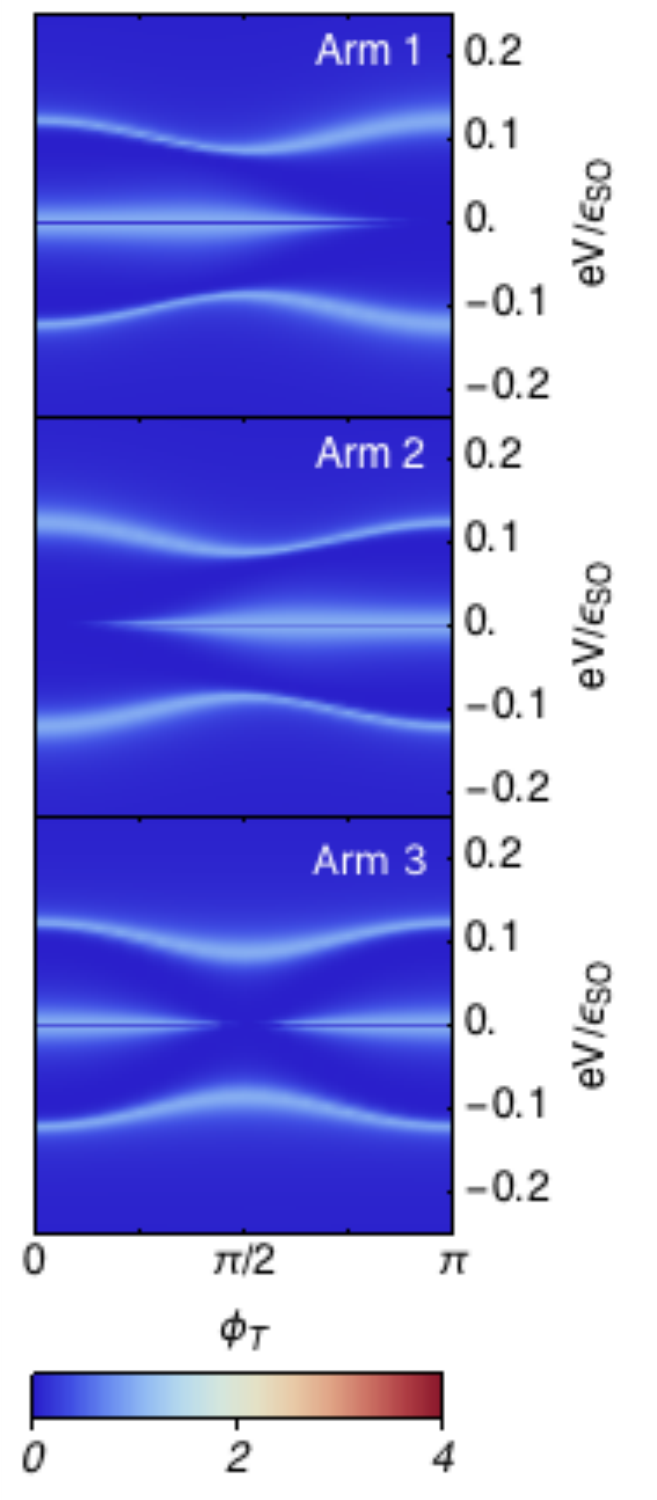}
\label{fig:TransferB}}
\caption{Differential tunneling conductance for the junction arms at distance $L_p=120$ nm from the origin and for fixed magnetic field $h=8$ $\epsilon_\text{SO}$. \protect\subref{fig:TransferA} Conductance as a function of bias voltage $V$ and phase variation $\phi_T$: $\phi_{s1}=\pi$, $\phi_{s2}=\pi$, $\phi_{s3}=\pi/2+\phi_T$.  \protect\subref{fig:TransferB} Conductance as a function of $V$ and $\phi_T$:  $\phi_{s1}=\pi-\phi_T$, $\phi_{s2}=\pi+\phi_T$, $\phi_{s3}=\pi/2$.}
\label{fig:Transfers}
\end{figure}

In Fig. \ref{fig:Transfers} we present results indicating such EMZM transfers. In Fig. \ref{fig:TransferA} we have set $\phi_{s1}=\pi$, $\phi_{s2}=\pi$, and varied $\phi_{s3}=\pi/2+\phi_T$ by tuning $\phi_T$ from $0$ to $\pi$. For $\phi_T=0$, ZBPs can be seen in arms $1$ and $3$ while for $\phi_T = \pi$ the ZBPs are in arms $2$ and $3$. 

In Fig. \ref{fig:TransferB}, $\phi_{s1}=\pi-\phi_T$, $\phi_{s2}=\pi+\phi_T$, and $\phi_{s3}=\pi/2$. Again, for $\phi_T=0$ there are ZBPs in arms $1$ and $3$ while for $\phi_T=\pi$ the peaks are in arms $2$ and $3$.

 We see that the effect of tuning the phases in both cases is to ``move'' a ZBP from arm $1$ to arm $2$ while a ZBP remains in arm $3$. However, in the latter arm, the conductance spectra differ between Figs. \ref{fig:TransferA} and \ref{fig:TransferB} in the sense that in the latter figure, the ZBP disappears temporarily for $\phi_T=\pi/2$ implying a disappearance of the EMZM in arm $3$. This feature can be explained in the following way. For the type of transfer in Fig. \ref{fig:TransferB}, when tuning two phases, there is a point in the phase parameter space where the phases in arms $1$ and $2$ coincide during the tuning. At precisely that point, arm $3$ is phase shifted with respect to the other two arms. The EMZM is then located only in arms $1$ and $2$ but its weight is redistributed to arm $3$ again when the phases are shifted further. This coincidence of phases does not occur in Fig. \ref{fig:TransferA} and the ZBP is present in arm $3$ during the whole process. We checked that for other phase choices, transfers between other arms are possible and yield similar results. 

We have shown that EMZMs can be transferred between arms in the junction and that this process can be detected in low bias tunneling conductance experiments. In particular, when the EMZM is located in two of the three arms, the tunneling conductance in the third arm is zero for a finite range of phases. See for instance the middle panels of Figs.~\ref{fig:TransferA} and \ref{fig:TransferB}, for low values of the phase.

Finally, we mention that we have checked that symmetry respecting weak disorder does not change the qualitative features of our results. Importantly, to successfully transfer the EMZMs between arms, no PTRS breaking disorder can be present. 

\section{Experimental Aspects}
\label{sec:Experimental Aspects}
In this section, we briefly comment on the experimental aspects of our setup. We believe that with present experimental techniques, the presented T-junction should be accessible. Nano-fabrication of proximity induced wires with connected leads has been reported by several groups and setups with connected wires have also been realized \cite{Nanocross}.

Varying the SC phases of the different arms can be achieved by connecting the outer regions of underlying $s$-wave SCs such that two loops are formed. If the areas in these loops are different, an external and tunable magnetic field will vary two phase differences between the SCs differently. The tunneling conductances in the NM region can then be measured individually as the magnetic flux is varied.

Regarding the tunneling probes, we have treated them as completely independent. In our calculations, we assumed that only one probe is active at any given time. Switching the probes on and off should not pose any experimental problems. 

As previously discussed, the behaviour of the EMZMs is insensitive to microscopical details, such as disorder, and how the three wires are connected. This also holds for the interfaces between the central region segments and the topological SC segments. However, it may be beneficial to have a weak coupling between the wires to reduce the overlap between EMZMs in the normal region and the outer edge MZMs.  Moreover, since the zero modes are extended uniformly over quite large regions in the wires, the exact location of tunneling probes is not very important, in contrast to probing local MZMs. 

In light of the discussion in Sec. \ref{sec:Existence and location of EMZMs}, the results presented here provide a signal of induced $p$-wave superconductivity, which in the class of proximity induced nanowires is highly desirable, and can be experimentally tested. 

\section{Conclusions}
\label{sec:Conclusions}
In this paper, we studied a topological superconducting-normal metal T-junction. We found that this system naturally hosts zero energy Andreev bound states which are of self-conjugate Majorana nature. These ``extended Majorana zero modes'' were shown to originate from perfect Andreev reflection upon the topological superconductors and also to be spatially extended with a uniform density over quite large regions ($\approx 100$ nm) in our model of the junction. Most importantly, if the junction respects pseudo time-reversal symmetry, we showed that the EMZMs distribute themselves only in two out of the three arms in the junction and that control of the superconducting phases allows for transfer of an EMZM between the junction arms. The location of the EMZMs can be probed by tunneling spectroscopy. Since we considered the long junction limit, the extended nature of the zero modes is crucial for our results.

We did not consider the braiding of localized MZMs in T-junctions (as explained in detail in Ref.~\onlinecite{hyart-qc}), but rather concentrated on the properties of EMZMs, and how they can be used to probe topological superconductors.

We supported our findings by a numerical tight-binding model of topologically superconducting nanowires and demonstrated that our results should be experimentally accessible with tunneling spectroscopy.

Since our results are highly dependent on the effective $p$-wave nature of the superconducting wires, we hope that our findings can motivate further experiments to reveal new insights in the field of topological superconductivity and Majorana physics. 

\section*{ACKNOWLEDGMENTS}
We gratefully acknowledge Thors Hans Hansson for numerous enlightening discussions and suggestions. We also thank Stefan Rex for helpful comments on the manuscript. C.S. thanks Annica Black-Schaffer, Göran Johansson, Mikael Fogelström and Iman Mahyaeh for helpful discussions and comments. This research was sponsored, in part, by the Swedish research council.

\appendix
\section{Symmetries of the reflection matrix and conductance quantization}
\label{sec:Symmetries of the reflection matrix}
It has been shown that particle-hole symmetry (PHS) strongly restricts the reflection matrix of an interface between a topological SC and a NM lead \cite{BrouwerZeroBias,Lee2009,FulgaQuantumNumber,AltlandZeroBias}. This restriction leads to a topological transport signature in terms of a quantized ZBP in the tunneling conductance. In this Appendix, we briefly review the derivation of these results.

For a single lead connected to a large SC (such that charging effects are negligible), and for energies much smaller than the gap, $\epsilon\ll |\Delta|$, the scattering matrix is a reflection matrix, $r(\epsilon)$, relating outgoing to incoming states by
\begin{equation}
\label{eq:reflectionMatrix1}
\Psi_\text{out} = r(\epsilon) \Psi_\text{in},
\end{equation}
where $\Psi_\text{in/out}$ are vectors containing the amplitudes of scattering states with incoming and outgoing momenta respectively. For an accessible introduction to mesoscopic scattering theory, see Ref. \onlinecite{DattaBook}.
 In the particle-hole basis, the reflection matrix is most conveniently divided into sub-blocks
\begin{equation}
\label{eq:reflectionMatrix2}
 r(\epsilon) =  \begin{pmatrix}
r_{ee}(\epsilon) & r_{eh}(\epsilon) \\
r_{he}(\epsilon) & r_{hh}(\epsilon)
\end{pmatrix}
\end{equation}
where $r_{ee}$ is reflection amplitude for an incoming electron, $r_{hh}$ is the reflection amplitude for an incoming hole and $r_{eh}$ and $r_{he}$ are Andreev reflection amplitudes which converts incoming electrons to outgoing holes and vice versa. The tunneling conductance for zero temperature and small bias voltages $V$ is dominated by Andreev processes and is given by \cite{blonder1982transition}
\begin{equation}
\label{eq:ConductanceFormula}
G(V) = \frac{2e^2}{h}\text{Tr}\left(r_{eh}^{} (eV) r_{eh}^\dag(eV)\right),
\end{equation}
where it is assumed that there is neither single particle transmission into the SC nor to other leads. The trace is taken over the channels in the lead.

Considering a spin-less single channel lead attached to a spin-less SC, we first note that the reflection matrix blocks are scalars. Secondly, the reflection matrix is unitary due to probability flux conservation and PHS enforces the zero energy constraint $\tau_x r^{}(0)\tau_x=r^*(0)$. These two restrictions allow only two possibilities for the reflection matrix entries: either $|r_{ee}(0)|=1$, $| r_{eh}(0)|=0$ or $|r_{ee}(0)|=0$, $|r_{eh}(0)|=1$ which have been shown to correspond to the trivial and topological regimes of the SC respectively. PHS implies additionally $r^{}_{eh}(0)=r^*_{he}(0)$. It follows from Eq. \eqref{eq:ConductanceFormula} that the zero bias conductance is quantized to $2e^2/h$ in the topological phase. It has been shown that this result persists even for certain types of interactions and also for spin-full leads \cite{fidkowski2012universal}.


\end{document}